\documentclass[12pt]{article}
\usepackage{axodraw}

\newcommand{\bea}{\begin{eqnarray}}
\newcommand{\eea}{\end{eqnarray}}

\title{{\small\hfill IMSc/2006/01/01}\\
\textbf{Instability in scalar channel of fermion-antifermion scattering amplitude
in massless QED$_3$ and anomalous dimensions of composite operators}}

\author{Indrajit Mitra$^{a}$\footnote{E-mail: indrajit.mitra@saha.ac.in},
Raghunath Ratabole$^{b}$\footnote{E-mail: raghu@imsc.res.in; present address:
             Birla Institute of Technology and Science, Pilani, Goa Campus,
             Zuarinagar 403726, Goa, India}~ and
H. S. Sharatchandra$^{b}$\footnote{E-mail: sharat@imsc.res.in} \\\\
$^a$ Theory Group, Saha Institute of Nuclear Physics, 1/AF Bidhan-Nagar,\\
Kolkata 700064, India\\
$^b$ The Institute of Mathematical Sciences, C.I.T. Campus, Taramani P.O.,\\
Chennai 600113, India}
\date{}
\begin{document}
\maketitle
\begin{abstract}

Instability in the scalar channel of the fermion-antifermion scattering amplitude
in massless QED$_3$ for number of flavours less than the critical value $128/3\pi^2$
is demonstrated. The anomalous dimensions of gauge-invariant composite operators
are determined to $O(1/N)$. Exponentiation of the
$O(1/N)$ infrared logarithm is explicitly demonstrated by evaluating
the contribution of the ladder diagrams.

\end{abstract}
\newpage
%%%%%%%%%%%%%%%%%%%%%%%%%
%%%%%%%%%%%%%%%%%%%%%%%%%

Massless QED$_3$ provides a theoretical laboratory for studying infrared (IR) 
divergences,
and for investigating dynamical breaking of chiral symmetry 
\cite{jackiw1, jackiw2, jackiw3, chisb1, chisb2, chisb3, chisb4, chisb5}.
It has also emerged as a leading paradigm for understanding the pseudogap phase
of cuprate superconductors \cite{rantner1, rantner2}. Lattice simulations of this theory
also form an active area of research \cite{kogut1, kogut2}. Recently, we have obtained
a rather detailed understanding of the IR behaviour of this theory
\cite{mrs1, mrs2, mrs3}. For a particular value of the gauge parameter in a 
non-local gauge 
\cite{nonlocal1, nonlocal2, nonlocal3, nonlocal4, nonlocal5}, the IR behaviour is of a conformal field theory
with canonical dimension for the fermion but a scale dimension one for the photon \cite{mrs1}.
This value of the gauge parameter is fixed at each order in $1/N$ ($N$ being the number of
fermion flavours). For other values of the gauge parameter in this non-local gauge,
the IR behaviour is again a conformal field theory, but with an anomalous dimension
of the fermion which is linear in the gauge parameter, and can take arbitrary values.
In the usual local gauge, the fermion correlation functions are exponentially
damped in the IR, as if a mass gap is present. However, this is only a gauge artifact
and illustrates the pitfalls involved in summing the severe IR divergences of the 
perturbation theory \cite{mrs2}. 

This emphasizes the need to study the IR behaviour of gauge-invariant correlation
functions, as they correspond to physical observables. The simplest such objects are
the Green functions involving only photons. They have a power law behaviour in the IR,
as per a scale dimension two (in contrast to the canonical dimension $3/2$) for the
field strength $F_{\mu\nu}(x)$. Another class of objects, which can be related to 
experiments which probe single fermions, are the correlations of gauge-invariant dressed 
fermions. We have argued in Ref.\ \cite{mrs3} that only the isotropic (in Euclidean
space-time) dressing is relevant, and with this dressing, the fermion correlations
have again a power law behaviour but with a negative anomalous dimension.

There is another class of correlations that are directly relevant to the experiments.
These are the response functions of QED$_3$ \cite{franz} corresponding to correlations of gauge-invariant
composite operators of the form $\bar\psi(x)\Gamma\psi(x)$, where the matrix
$\Gamma$ is any element of the algebra of gamma-matrices.
We address their IR behaviour in this paper.
We will first perform an $O(1/N)$ calculation which will give an IR logarithm, heralding
an anomalous dimension for the composite operator.
We will then explicitly 
demonstrate how the ladder diagrams lead to an exponentiation of this IR logarithm.\footnote{The anomalous 
dimensions for the cases $\Gamma=1$ and $\Gamma=\gamma_5$ in the ladder approximation
were also addressed in Ref.\ \cite{ghr}.
The result is discussed Ref.\ \cite{gkr}.}

A crucial issue in QED$_3$ is  dynamical breaking of chiral symmetry and the critical number of flavours
below which this takes place 
\cite{chisb1, chisb2, chisb3, chisb4, chisb5}.
It is always very difficult to calculate the condensate $<\bar\psi\psi>$. There are different
claims for the critical number of flavours $N_c$,
ranging from $N_c=1$ to $N_c=4$.\footnote{For a review, see, for example, Ref.\ \cite{review}.}
In the course of our analysis, we calculate the fermion-antifermion scattering amplitude\footnote{The 
fermion-antifermion scattering amplitude was also considered in Ref.\ \cite{atw}.}
in the scalar channel for vanishing total energy-momentum. We demonstrate instability in
the vacuum in which this scattering amplitude is calculated, for number of flavours less
than the critical value $N_c=128/3\pi^2$. 
Our calculation is important for the following reasons.
Firstly, {\it we need only ladder diagrams} and not self-consistent solution of the gap equation.
Secondly, {\it we relate the instability to a robust physical mechanism}: an attractive inverse-square
potential (with momentum variables in the place of coordinate variables) arising from the masslessness 
of the fermions and the feature that the small-momentum behaviour of the photon propagator
is inversely linear.

%%%%%%%%%%%%%%%%%%%%%%%%%%%%%%%%%%%%%%%%%%%%%%%%%%%%%%%%%%
%Correlation
%
\begin{figure}
\begin{center}
%\fcolorbox{white}{white}{
  \begin{picture}(331,142) (30,-14)
    \SetWidth{0.5}
%    \SetColor{Black}
    \Line(30,69)(60,69)\Line(30,65)(60,65)%%JaxoDrawID:DoubleLine(2)
    \GOval(106,67)(40,15)(0){0.882}
    \Line(60,69)(60,65)
    \Line(60,67)(92,88)
    \ArrowLine(113,102)(150,127)
    \ArrowLine(150,7)(113,33)
    \Line(60,67)(93,46)
    \Text(151,18)[lb]{\Large{{$l$}}}
    \Text(41,74)[lb]{\Large{{$q$}}}
    \Text(143,102)[lb]{\Large{{$l+q$}}}
    \Line(240,69)(240,65)
    \ArrowLine(240,67)(330,127)
    \ArrowLine(330,7)(240,67)
    \Photon(301,109)(301,26){5}{5.5}
    \Text(221,75)[lb]{\Large{{$q$}}}
    \Text(323,103)[lb]{\Large{{$l+q$}}}
    \Text(331,18)[lb]{\Large{{$l$}}}
    \Text(77,-15)[lb]{\Large{{$(A)$}}}
    \Text(256,-15)[lb]{\Large{{$(B)$}}}
    \Line(210,69)(240,69)\Line(210,65)(240,65)%%JaxoDrawID:DoubleLine(2)
  \end{picture}
%}
%%%%%%%%%%%%%%%%%%%%%%%%%%%%%%%%%%%%%%%%%%%%%%%%%%%%%%%%%%%%%%%
\caption[]{\sf (A) The correlation function 
$<\bar\psi(x)\Gamma\psi(x)\psi_\alpha(y)\bar\psi_\beta(z)>$  in momentum space
(the figure is for $\Gamma=1$). (B) The $O(1/N)$ contribution.
\label{f:corr}}
%%%%%%%%%%%%%%%%%%%%%%%%%%%%%%%%%%%%%%%%%%%%%%%%%%%%%%%%%%%%%%%%%%
\end{center}
\end{figure}
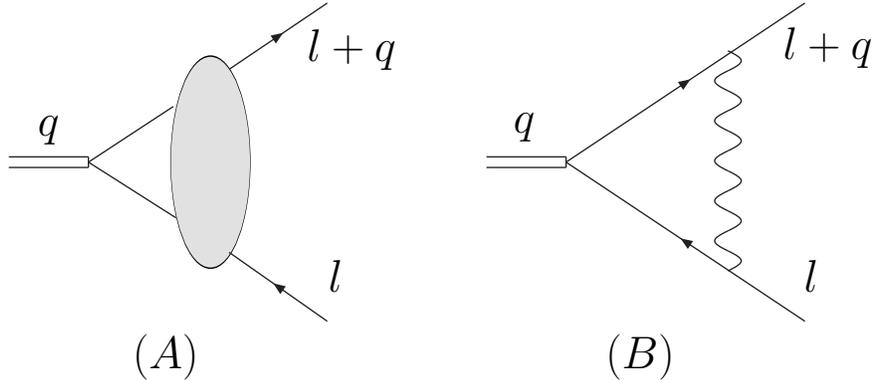
%%%%%%%%%%%%%%%%%%%%%%%%%%%%%%%%%%%%%%%%%%%%%%%%%%%%%%%%%%%%%%%%%%%%

Let us consider (the Fourier transform of) the correlation function
$<\bar\psi(x)\Gamma\psi(x)\psi_\alpha(y)\bar\psi_\beta(z)>$ (see Fig.\ \ref{f:corr}A).
This correlation is gauge-dependent, but from it, one can extract the (gauge-invariant)
anomalous dimension of the composite operator $\bar\psi\Gamma\psi$, as follows.

The $O(1/N)$ contribution to this correlation function, as shown in Fig.\ \ref{f:corr}B, 
is given by
\bea
\int\frac{d^3k}{(2\pi)^3}\,\,e\gamma_\sigma
         \frac{1}{\rlap/l +\rlap/q+\rlap/k}\, \Gamma
	 \frac{1}{\rlap/l+\rlap/k}\,e\gamma_\rho
	 \frac{\delta_{\sigma\rho}-\xi k_\sigma k_\rho/k^2}{k^2+\mu k}
\eea
with $\mu=Ne^2/8$ (for $N$ four-component spinors).
Here we have used the photon propagator \cite{mrs1} in the $1/N$ expansion
(in which there is a resummation of chains of one-loop vacuum polarization
diagrams on every photon propagator) with a non-local gauge-fixing term.\footnote{The gauge-fixing 
term is given by Eq.\ (25) of Ref.\ \cite{mrs2}. The gauge parameter
$\alpha$ of Ref. \cite{mrs2} is related to the gauge parameter $\xi$ of the present paper
and of Ref.\ \cite{mrs1} by $\alpha=1-\xi$.}
This non-local gauge ensures that the IR behaviour of the propagator is inversely linear in
momentum for arbitrary value of the gauge parameter $\xi$. The 
value of the gauge parameter can then be fixed to each order in $1/N$ such that the IR 
logarithms are absent in the fermion self-energy and in other Green functions of 
elementary fields \cite{mrs1}. This choice of $\xi$ will simplify the ladder diagrams,
to be considered later in this paper.

We will now follow the calculation of Appendix A of Ref.\ \cite{mrs1}.
Choose
$l_\mu=\rho L_\mu$ and $q_\mu=\rho Q_\mu$. Also let $k_\mu=\rho K_\mu$. Then
the integral equals
\bea
e^2\int\frac{d^3K}{(2\pi)^3}\,\,\gamma_\sigma
         \frac{1}{\rlap/L +\rlap/Q+\rlap/K}\, \Gamma
         \frac{1}{\rlap/L+\rlap/K}\,\gamma_\rho
         \frac{\delta_{\sigma\rho}-\xi K_\sigma K_\rho/K^2}{\rho K^2+\mu K}\,.   \label{gama}
\eea
For $\rho\rightarrow 0$, with $L_\mu$ and $ Q_\mu$ of $O(1)$, this is IR finite but logarithmically 
ultraviolet (UV) divergent.
Letting the
divergent part be $C\ln\rho$, the coefficient $C$ can be obtained by the action of
$[\rho(d/d\rho)]_{\rho=0}$ on the last expression 
(it is convenient to replace $K$ by $k/\rho$ before setting $\rho=0$):
\bea
C=-e^2\int\frac{d^3k}{(2\pi)^3}\,\,\gamma_\sigma \frac{\rlap/k}{k^2}
        \, \Gamma \frac{\rlap/k}{k^2}\,\gamma_\rho
        \frac{\delta_{\sigma\rho}-\xi k_\sigma k_\rho/k^2}{(k+\mu)^2}\,.
\eea
In the numerator, the part proportional to $\xi$ equals $-\xi k^2\Gamma$. Writing the other part
as $k_\mu k_\nu \gamma_\sigma\gamma_\mu\Gamma\gamma_\nu\gamma_\sigma$, we replace $k_\mu k_\nu$
by $\frac{1}{3}k^2\delta_{\mu\nu}$. Now consider two types of vertices \cite{franz}:
\bea 
\gamma_\sigma\gamma_\mu\Gamma\gamma_\mu\gamma_\sigma=\kappa\Gamma
\eea
with  $\kappa=9$ if $\Gamma$ commutes or anticommutes with all the three gamma-matrices, and $\kappa=1$
if $\Gamma$ anticommutes with one or two of the gamma-matrices and commutes with the rest. Thus the
numerator equals $k^2(\frac{1}{3}\kappa-\xi)\Gamma$, and on doing the angular integration,
\bea
C=-\frac{4}{\pi^2 N}\Bigg(\frac{1}{3}\kappa-\xi\Bigg)\Gamma\,.
\eea
Now, if we choose the value $\xi=1/3$ for which the $O(1/N)$ logarithmic divergence in the fermion
self-energy is absent \cite{mrs1}, the coefficient of $\Gamma$ in $C$ gives the anomalous dimension of the
composite operator $\bar\psi\Gamma\psi$.\footnote{This follows from the application of 
the Callan-Symanzik equation
to the correlation function under consideration; see Ref.\ \cite{pes}
for a similar situation.}
Thus for the vertices with $\kappa=9$, the composite operator (for example, $\bar\psi\psi$)
has an anomalous dimension
\bea
\eta=-\frac{32}{3\pi^2 N}\,,                                  \label{eq:eta}
\eea
while for the vertices with $\kappa=1$, the composite operator does not have an anomalous dimension.
Our result agrees with that of Ref.\ \cite{franz}. The apparent discrepancy 
of a factor of $-2$ between Eq. (\ref{eq:eta}) and
the result of Ref.\ \cite{franz} is merely due to difference in the definition of anomalous dimension.
In particular, the authors of Ref.\ \cite{franz} consider the dependence of 
$<\bar\psi(x)\Gamma\psi(x)\bar\psi(y)\Gamma\psi(y)>$ on the external momentum, and this correlation
has $\it two$ factors of the wave-function renormalization $Z_{\bar\psi\Gamma\psi}$ instead of $one$ factor
as in our correlation. It is to be noted that we have obtained the anomalous
dimension by performing a one-loop calculation,
instead of the two-loop calculation of Ref.\ \cite{franz}.

%%%%%%%%%%%%%%%%%%%%%%%%%%%%%%%%%%%%%%%%%%%%%%%%%%%%%%%%%%%%%%%%%%%%%%%%%%%%%%%%%
%Ladder diagrams
%
\begin{figure}
\begin{center}
%\fcolorbox{white}{white}{
  \begin{picture}(481,90) (45,-30)
  \SetOffset(70,0)
    \SetWidth{0.6}
%    \SetColor{Black}
    \SetScale{0.7}
    \Line(45,17)(90,17)\Line(45,13)(90,13)%%JaxoDrawID:DoubleLine(2)
    \SetWidth{0.7}
    \Line(90,15)(150,60)
    \Line(90,15)(150,-30)
    \Line(165,15)(180,15)
    \Line(173,22)(173,9)
    \Line(240,15)(300,60)
    \Line(240,15)(300,-30)
    \Photon(276,42)(278,-14){5}{4.5}
    \Line(315,15)(330,15)
    \Line(323,22)(323,9)
    \SetWidth{0.6}
    \Line(195,17)(240,17)\Line(195,13)(240,13)%%JaxoDrawID:DoubleLine(2)
    \Line(345,17)(390,17)\Line(345,13)(390,13)%%JaxoDrawID:DoubleLine(2)
    \SetWidth{0.5}
    \Line(390,15)(450,60)
    \Line(390,15)(450,-30)
    \Photon(419,35)(419,-8){5}{3.5}
    \Photon(436,50)(437,-21){5}{5.5}
    \Line(465,15)(480,15)
    \Line(465,15)(480,15)
    \Line(473,22)(473,9)
    \Vertex(495,15){1.41}
    \Vertex(510,15){1.41}
    \Vertex(525,15){1.41}
    \Line(90,17)(90,13)
    \Line(240,17)(240,13)
    \Line(390,17)(390,13)
  \end{picture}
%}
%%%%%%%%%%%%%%%%%%%%%%%%%%%%%%%%%%%%%%%%%%%%%%%%%%%%%%%%%%%%%%%%%
\caption[]{\sf Ladder diagrams contributing to the correlation function.
\label{f:ladder}}
%%%%%%%%%%%%%%%%%%%%%%%%%%%%%%%%%%%%%%%%%%%%%%%%%%%%%%%%%%%%%%%%%%
\end{center}
\end{figure}
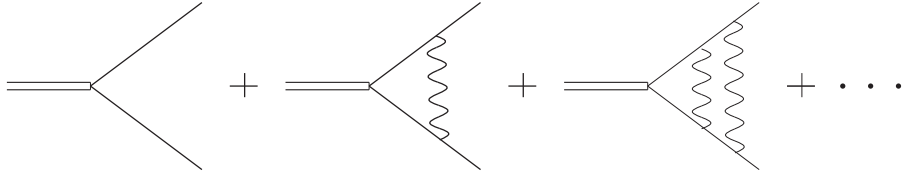
%%%%%%%%%%%%%%%%%%%%%%%%%%%%%%%%%%%%%%%%%%%%%%%%%%%%%%%%%%%%%%%%%%%%

The anomalous dimension was heralded by the appearance of an IR ($\rho\rightarrow 0$) logarithm in external
momenta. In order to address the exponentiation of this logarithm, we consider the contribution of the ladder 
diagrams (see Fig.\ \ref{f:ladder}). 
We choose to work with the special gauge parameter $\xi_0$ of our non-local gauge
for which the logarithmic terms are absent at all orders in the fermion self-energy and in other Green functions
of the elementary fields $\psi$, $\bar\psi$ and $A_\mu$. Then the full fermion propagator has the behaviour of
the free propagator $1/\rlap/k$, and the full vertex $e\gamma_\mu$, when all the momenta are small. Therefore,
{\it keeping only the bare fermion propagators and the bare vertices in Fig.\ \ref{f:ladder} is justified.}
Moreover, we will demonstrate the exponentiation of only the lowest order contribution to the anomalous dimension.
So we will not consider the diagrams with crossed photons, which will contribute at higher orders in $1/N$.

%%%%%%%%%%%%%%%%%%%%%%%%%%%%%%%%%%%%%%%%%%%%%%%%%%%%%%%%%%%%%%%%%%%%%%%
%n-loop ladder
%
\begin{figure}
\begin{center}
%\fcolorbox{white}{white}{
  \begin{picture}(375,198) (30,-4)
    \SetWidth{0.5}
%    \SetColor{Black}
    \Text(32,97)[lb]{\Large{{$q=0$}}}
    \Text(155,88)[lb]{\Large{{$l_1-l_2$}}}
    \SetWidth{0.5}
    \Line(30,95)(75,95)\Line(30,91)(75,91)%%JaxoDrawID:DoubleLine(2)
    \Line(75,93)(405,183)
    \Line(75,93)(405,3)
    \Photon(148,113)(148,73){5}{3.5}
    \Photon(210,130)(210,56){5}{5.5}
    \Photon(329,161)(329,24){5}{10.5}
    \Text(338,87)[lb]{\Large{{$l_n-l$}}}
    \Vertex(255,93){1.41}
    \Vertex(270,93){1.41}
    \Vertex(285,93){1.41}
    \Vertex(240,93){1.41}
    \Vertex(300,93){1.41}
    \Text(97,109)[lb]{\Large{{$l_1$}}}
    \Text(100,67)[lb]{\Large{{$l_1$}}}
    \Text(162,123)[lb]{\Large{{$l_2$}}}
    \Text(162,51)[lb]{\Large{{$l_2$}}}
    \Text(361,178)[lb]{\Large{{$l$}}}
    \Text(364,-4)[lb]{\Large{{$l$}}}
    \Text(291,160)[lb]{\Large{{$l_n$}}}
    \Text(293,13)[lb]{\Large{{$l_n$}}}
    \Line(75,95)(75,92)
  \end{picture}
%}:w

%%%%%%%%%%%%%%%%%%%%%%%%%%%%%%%%%%%%%%%%%%%%%%%%%%%%%%%%%%%%%%%%%
\caption[]{\sf The $n$-loop ladder diagram: the spinor algebra is
drastically simplified for $q=0$.
\label{f:nloop}}
%%%%%%%%%%%%%%%%%%%%%%%%%%%%%%%%%%%%%%%%%%%%%%%%%%%%%%%%%%%%%%%%%%
\end{center}
\end{figure}
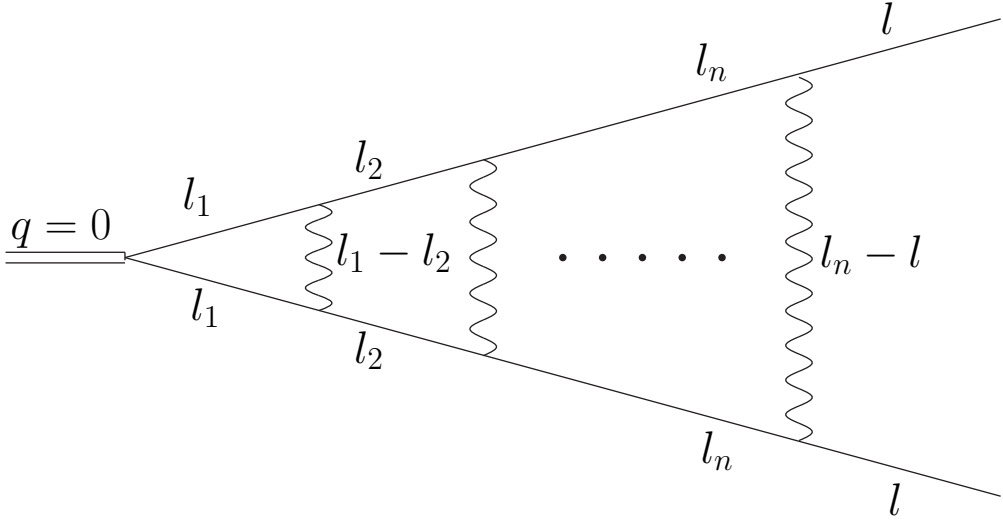
%%%%%%%%%%%%%%%%%%%%%%%%%%%%%%%%%%%%%%%%%%%%%%%%%%%%%%%%%%%%%%%%%%%%

Power counting of the subdiagrams shows that there are no UV divergences. We are interested in the
IR behaviour. We will also consider the case $q=0$. To extract the anomalous dimension of the composite 
operator, we need the correlation function at non-exceptional and Euclidean external 
momenta.
There is a
possibility of additional IR divergences at exceptional values.
However, we consider this case as there is an enormous simplification in the spinor algebra, so that the
power law in the IR comes out in a simple and straightforward way.

We take $\Gamma$ to commute or anticommute with all the gamma-matrices, which is the case for which we obtained
a log divergence in our one-loop calculation. Now from the $l_1$-loop in Fig.\ \ref{f:nloop}, we get the spinor factors
\bea
e^2\gamma_\mu\rlap/l_1\rlap/l_1\gamma_\nu\big(\delta_{\mu\nu}-\xi_0
  (l_1-l_2)_\mu (l_1-l_2)_\nu/(l_1-l_2)^2\big)=e^2(3-\xi_0)l_1^2\,.
\eea
As four gamma-matrices are involved, this contribution is insensitive to our sign convention
and the choice of the action. Similar simplification occurs for each of the loops of Fig.\ \ref{f:nloop}. Thus,
all spinor dependence is absent, as if the fermion is replaced with a scalar. We get
a contribution
\bea
(e^2(3-\xi_0))^{n}
\int\frac{d^3l_1}{(2\pi)^3 l_1^2}\,
\int\frac{d^3l_2}{(2\pi)^3 l_2^2}\,\,\cdots
\int\frac{d^3l_n}{(2\pi)^3 l_n^2}\,\,
\frac{1}{\mu|\vec l_1-\vec l_2|}\,
\frac{1}{\mu|\vec l_2-\vec l_3|}\,\cdots\,
\frac{1}{\mu|\vec l_n-\vec l|}\,,
\eea
from $n$ ladders,
where we have kept the leading IR behaviour of the photon propagator, and presumed a cut-off
in the UV for each integration. Integration over $l_1$ leads to $\sim\ln l_2$
for small $l_2$ (this is because $l_2$ serves as the IR cutoff for an otherwise log divergent
$l_1$ integration). This when fed into the $l_2$ integration, gives $\sim d(\ln l_2)\ln l_2$
with an IR cutoff $l_3$, which is $\sim\frac{1}{2!}(\ln l_3)^2$. Continuing in this way,
the $n$-loop integration of
Fig.\ \ref{f:nloop} 
gives a contribution $\sim\frac{1}{n!}(\ln l)^n$. A sum over the number of ladders $n$ , 
as in Fig.\ \ref{f:ladder},
then gives a power in the fermion momentum $l$, namely, $\exp[c(e^2/\mu)(3-\xi_0)\ln l]
=l^{c(e^2/\mu)(3-\xi_0)}$, where $c$ is a numerical factor. This argument thus suggests an 
anomalous dimension for composite operators like $\bar\psi\psi$.

It should be noted that it is the logarithm from the $l_1$-loop which eventually gets
exponentiated. The vertex in the $l_1$-loop is not a $\bar\psi\rlap/A\psi$ vertex,
and so our proof \cite{mrs1} of absence of anomalous dimension for Green functions involving
{\it elementary} fields in the gauge $\xi=\xi_0$ does not apply for this correlation function. 

%%%%%%%%%%%%%%%%%%%%%%%%%%%%%%%%%%%%%%%%%%%%%%%%%%%%%%%%%%%%%%%%%%
%Fermion-antifermion kernel
%
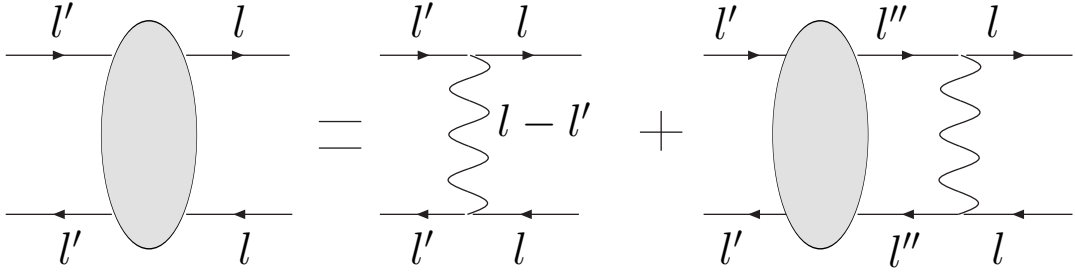
\begin{figure}
\begin{center}
%\fcolorbox{white}{white}{
  \begin{picture}(403,103) (37,-52)
    \SetWidth{0.5}
%    \SetColor{Black}
    \GOval(91,-2)(43,18)(0){0.882}
    \ArrowLine(338,-32)(300,-32)
    \ArrowLine(77,-32)(37,-32)
    \ArrowLine(105,28)(144,28)
    \ArrowLine(145,-32)(105,-32)
    \Line(155,3)(171,3)
    \Line(155,-7)(171,-7)
    \Photon(212,28)(211,-32){7.5}{3.5}
    \ArrowLine(214,28)(254,28)
    \ArrowLine(253,-32)(213,-32)
    \ArrowLine(210,-32)(178,-32)
    \ArrowLine(178,28)(211,28)
    \ArrowLine(300,28)(338,28)
    \GOval(344,-2)(43,18)(0){0.882}
    \ArrowLine(358,28)(396,28)
    \ArrowLine(396,-32)(358,-32)
    \Photon(397,28)(396,-32){7.5}{3.5}
    \ArrowLine(439,-32)(398,-32)
    \ArrowLine(399,28)(439,28)
    \ArrowLine(37,28)(77,28)
    \Text(223,-4)[lb]{\Large{{$l-l'$}}}
    \Text(54,35)[lb]{\Large{{$l'$}}}
    \Text(123,35)[lb]{\Large{{$l$}}}
    \Text(57,-51)[lb]{\Large{{$l'$}}}
    \Text(125,-51)[lb]{\Large{{$l$}}}
    \Text(190,35)[lb]{\Large{{$l'$}}}
    \Text(190,-51)[lb]{\Large{{$l'$}}}
    \Text(230,35)[lb]{\Large{{$l$}}}
    \Text(229,-51)[lb]{\Large{{$l$}}}
    \Text(305,34)[lb]{\Large{{$l'$}}}
    \Text(307,-51)[lb]{\Large{{$l'$}}}
    \Text(366,34)[lb]{\Large{{$l''$}}}
    \Text(371,-52)[lb]{\Large{{$l''$}}}
    \Text(408,35)[lb]{\Large{{$l$}}}
    \Text(410,-51)[lb]{\Large{{$l$}}}
    \Line(276,-1)(292,-1)
    \Line(284,7)(284,-8)
  \end{picture}
%}

%%%%%%%%%%%%%%%%%%%%%%%%%%%%%%%%%%%%%%%%%%%%%%%%%%%%%%%%%%%%%%%%%
\caption[]{\sf Integral equation for the fermion-antifermion scattering 
amplitude in the ladder approximation (and at vanishing total energy-momentum).
\label{f:kernel}}
%%%%%%%%%%%%%%%%%%%%%%%%%%%%%%%%%%%%%%%%%%%%%%%%%%%%%%%%%%%%%%%%%%
\end{center}
\end{figure}
%%%%%%%%%%%%%%%%%%%%%%%%%%%%%%%%%%%%%%%%%%%%%%%%%%%%%%%%%%%%%%%%%%%%

We now analyse this situation using an integral equation. Let us consider the kernel
$I(\vec l, \vec l')$ depicted in Fig.\ \ref{f:kernel}. By joining the $l'$ lines,
we can recover the correlation function considered earlier. We have
\bea
I(\vec l, \vec l')=\frac{1}{|\vec l-\vec l'|} +\lambda\int d^3 l''\frac{1}{l''^2}
                   \frac{I(\vec l'', \vec l')}{|\vec l- \vec l''|}     \label{eq:kernel}
\eea
where
\bea
\lambda=\frac{e^2}{(2\pi)^3\mu}(3-\xi_0)\,. 
\eea
Eq.\ (\ref{eq:kernel}) clearly holds for the case when the $l'$ lines are joined to give a 
$\bar\psi\psi$ vertex (that is, $\Gamma=1$), which carries zero spin.
Thus, Eq.\ (\ref{eq:kernel}) is for the scalar channel of the
fermion-antifermion scattering amplitude.  
We may convert this integral equation into a differential equation by
operating on it with $\nabla{_l^2}$:
\bea
\Bigg(\nabla{_l^2}+\frac{4\pi\lambda}{l^2}\Bigg)I(\vec l,\vec l')
                     =-4\pi\delta^{(3)}(\vec l-\vec l')\,.          \label{eq:diff}
\eea
This means that $I(\vec l, \vec l')$ is a propagator (Green function)
for the three-dimensional Schr\"odinger equation (with momentum variables in the place of
coordinate variables) with the potential
$V(\vec l)=-4\pi\lambda/l^2$. As $\lambda>0$ ($\xi_0$ being $1/3$) for us, this is
an attractive inverse square potential. This is a well-studied problem
in quantum mechanics \cite{landau}, and has unusual properties
due to exact scale invariance. We will return to this connection a little later.

The Green function can be determined using standard techniques \cite{jack}.
The ``radial" part $f_L(l,l')$ for the channel with angular momentum $L$ satisfies
\bea
\frac{1}{l}\frac{d^2}{dl^2}(lf_L)-\frac{L(L+1)-4\pi\lambda}{l^2}f_L
                     = -\frac{4\pi}{l^2}\delta(l-l')\,.                \label{eq:radial}
\eea
Note that the the $\lambda$-containing term (with $\lambda>0$) provides
a centripetal attraction in contrast to the centrifugal repulsion provided
by the angular momentum.
The radial equation (for $l\neq l'$) has the two power law solutions:
\bea
l^{-\frac{1}{2}\pm\alpha},\,\,\,\alpha=\sqrt{\Bigg(L+\frac{1}{2}\Bigg)^2-4\pi\lambda}\,.
\eea
%$l^{-\frac{1}{2}\pm\sqrt{(L+\frac{1}{2})^2-4\pi\lambda}}$. 
If $4\pi\lambda<1/4$
(which is true when the number of flavours $N>N_c=(16/\pi^2)(3-\xi_0)$),
$\alpha$ is real for each of $L=0,1,2,\cdots$. Now the
boundary conditions on $f_L(l,l')$ follows from the IR and UV finiteness of
of the integral on the right-hand side of Eq.\ (\ref{eq:kernel}):
$\lim_{l\rightarrow 0}l f_L(l,l')=0$ and
$\lim_{l\rightarrow \infty} f_L(l,l')=0$. Also, 
$f_L(l,l')= f_L(l',l)$. Therefore, we have $f_L(l,l')=Cl_<^{-1/2+\alpha} l_>^{-1/2-\alpha}$,
where
$l_<$ ($l_>$) is the smaller (larger) of $l$ and $l'$,
and $C$ is determined from the effect of the delta function in Eq.\ (\ref{eq:radial}). 
Thus, the solution for the Green function is obtained as
\bea
I(\vec l, \vec l')=\sum_{L=0}^\infty\sum_{M=-L}^L
                   \frac{2\pi}{\sqrt{(L+\frac{1}{2})^2-4\pi\lambda}}
                   \frac{1}{\sqrt{l_< l_>}}
                   \Bigg(\frac{l_<}{l_>}\Bigg)^{\sqrt{(L+\frac{1}{2})^2-4\pi\lambda}}
                   Y{_{LM}^*}(\theta',\phi')Y_{LM}(\theta,\phi)\,,                              \label{eq:sol}
\eea
where $(\theta,\phi)$ and $(\theta',\phi')$ are the angular variables for $\vec l$ and $\vec l'$
respectively.

For the quantum mechanical problem of a particle in an attractive inverse-square potential,
a coupling constant $4\pi\lambda$ exceeding the critical value
$4\pi\lambda_c=1/4$ leads to a singular situation: the particle falls to the centre \cite{landau}.
In our problem, this corresponds to $N<N_c=128/3\pi^2\approx 4.3$ for $\xi_0=1/3$. This is same as the
value for the critical number of flavours for dynamical chiral symmetry breaking, as obtained
from an analysis of gap equation \cite{nonlocal1, nonlocal5}. 

Let the eigenfunctions and eigenvalues of the Schr\"odinger operator $-\nabla{_l^2}-4\pi\lambda/l^2$
be denoted by $\chi_n(\vec l)$ and $\epsilon_n$ respectively. Then, from Eq.\ (\ref{eq:diff}),
\bea
I(\vec l, \vec l')=4\pi\sum_n \frac{\chi_n(\vec l)\chi{^*_n}(\vec l')}{\epsilon_n}\,.
\eea
There are no bound states (that is, $\epsilon_n < 0$) for $\lambda < \lambda_c$ \cite{landau, camblong},
but bound states appear at the critical coupling. $I(\vec l, \vec l')$ is the Bethe-Salpeter amplitude
in the channel with total spin zero and total energy-momentum zero. {\it This is possibly the simplest object
which brings out instability in the vacuum}. For total energy-momentum zero, only the mass-squared terms
in the poles in the amplitude survive. So having some $\epsilon_n < 0$ corresponds to a tachyon. 
This means an instability in the vacuum in which we have computed $I(\vec l, \vec l')$. The system cures
this by forming a condensate.

Let us now join the $l'$ lines in Fig.\ \ref{f:kernel} to obtain
the desired correlation function. Thus, we are to evaluate
\bea
{\cal I}(l)\equiv \int \frac{d^3 l'}{(2\pi)^3}\frac{1}{l'^2} e^2(3-\xi_0)\frac{1}{\mu}
     I(\vec l, \vec l')
\eea
in the limit $l\rightarrow 0$. Because of the integral over the angular variables for $\vec l'$,
only the $L=0$ term is picked out from $I(\vec l, \vec l')$. Then
\bea
{\cal I}(l)= \frac{e^2(3-\xi_0)}{4\pi^2\mu \sqrt{1/4-4\pi\lambda}}\int_0^\mu dl'
                   \frac{1}{\sqrt{l_< l_>}}
                   \Bigg(\frac{l_<}{l_>}\Bigg)^{\sqrt{1/4-4\pi\lambda}}\,.             \label{eq:X}
\eea
(The full $1/(k^2+\mu k)$ photon propagator provides an effective UV cutoff of $O(\mu)$.)
On splitting the range of integration into $(0,l)$ and $(l,\mu)$, and putting $l'=lx$,
we get the two integrals:
\bea
\int_0^1 dx\,x^{-1/2+\sqrt{1/4-4\pi \lambda}}+
\int_1^{\mu/l} dx\,x^{-1/2-\sqrt{1/4-4\pi \lambda}}\,.
\eea
While there is no divergence from the $x=0$ end, there is a divergence at $x= {\mu/l}$ for $l\rightarrow 0$.
On expanding $\sqrt{1/4-4\pi \lambda}$ in powers of $1/N$, it is found to equal $1/2+\eta$ at $O(1/N)$, where
$\eta$ is given in Eq.\ (\ref{eq:eta}). Also, the prefactor in the integral in Eq.\ (\ref{eq:X}) equals
$-\eta$, to the lowest order in $1/N$. Thus,
\bea
{\cal I}(l)=-\eta\int^{\mu/l} dx\,x^{-1-\eta}\sim\Bigg(\frac{l}{\mu}\Bigg)^\eta
\eea
for $l\rightarrow 0$. (Since we have a divergence for $l\rightarrow 0$, the contribution of unity
from the first diagram (the free-theory diagram) in Fig.\ \ref{f:ladder} can be neglected.) We have thus demonstrated
that the correlation function has a power law behaviour with the exponent $\eta$.\footnote{This agrees 
with the result of Ref.\ \cite{ghr} when the
appropriate limits are taken.} 

Let us now briefly consider the case
$q\neq 0$. Let the incoming
fermion-antifermion pair in the scattering amplitude
have the momenta $\vec l \pm \vec q/2$, and the outgoing pair
$\vec l' \pm \vec q/2$. Let us define $\vec L=\vec l/q$ and $\vec L'=\vec l'/q$, and scale the loop
momentum:
$\vec L''=\vec l''/q$. Also define
$J(\vec l, \vec l',\vec q)=q I(\vec l, \vec l',\vec q)$.
Then $J(\vec l, \vec l',\vec q)$ satisfies almost the same integral equation 
as Eq.\ (\ref{eq:kernel}) for  $I(\vec l, \vec l')$
with $\vec l\rightarrow \vec L$, $\vec l'\rightarrow \vec L'$, $\vec l''\rightarrow \vec L''$; only the factor
$1/l''^2$ in the last term 
gets replaced with $1/(|\vec L''+\hat q/2| |\vec L''-\hat q/2|)$, where $\hat q$ is the unit
vector along $\vec q$. (To simplify our analysis, we have used $1/k$ as the fermion
propagator.) This then gives the differential equation
\bea
\Bigg(\nabla{_L^2}+\frac{4\pi\lambda}{|\vec L+\hat q/2| |\vec L-\hat q/2|}\Bigg)J(\vec L,\vec L', \vec q)
                     =-4\pi\delta^{(3)}(\vec L-\vec L')\,.
\eea
So, for large $\vec L$, we still have an inverse square potential, but near $\vec L=\pm\hat q/2$
the potential is Coulomb-like.
A potential of this form has also been obtained\footnote{The authors of Ref.\ \cite{gusynin} 
find the same value
of $N_c$ in QED$_3$ as ours.
The
two-centre potential appears in Section 4 of Ref.\ \cite{isaev}.}
following other routes.

In this paper, we demonstrated 
instability in the scalar channel of the fermion-antifermion scattering amplitude
in massless QED$_3$, for number of flavours less than the critical value $128/3\pi^2$
for which spontaneous chiral symmetry breaking takes place. This was done using
only the ladder diagrams, and the instability was linked to the robust physical
mechanism of an attractive inverse-square potential. 
We also determined the anomalous dimensions of the gauge-invariant composite operators
to $O(1/N)$, first as coefficient of IR logarithm, and then as exponent of
power law.

\section*{Acknowledgements}

We would like to thank Z. Tesanovic for stimulating correspondence. We also thank
A. Bashir and V. Gusynin for useful communications.
I.M. thanks IMSc, Chennai for hospitality during the course of this work.
The LaTeX codes for the figures in this paper were generated primarily using
JaxoDraw \cite{jaxo}.

\end{document}